\documentclass[floatfix,twocolumn
]{revtex4}

\usepackage{amsmath}
\usepackage{graphicx}

\def\GeV{{\rm\ GeV}}
\def\ve{\varepsilon}
\def\CG{{\cal G}}
\def\be{\begin{equation}}
\def\ee{\end{equation}}
\def\bea{\begin{eqnarray}}
\def\eea{\end{eqnarray}}

\def\section#1{\par{\it #1.}}

\begin{document}

\title{Two-photon exchange amplitudes for the elastic $ep$ scattering
at $Q^2=2.5\GeV^2$\\ from the experimental data}

\author{Dmitry~Borisyuk}
\author{Alexander~Kobushkin}

\affiliation{Bogolyubov Institute for Theoretical Physics,
Metrologicheskaya street 14-B, 03680, Kiev, Ukraine}

\begin{abstract}
We extract two-photon exchange amplitudes 
for the elastic electron-proton scattering at $Q^2=2.5\GeV^2$
from the unpolarized cross-section
and recent polarization transfer measurements.
There are three independent amplitudes, but only one of them,
$\delta\CG_M$, can be determined with a reasonable accuracy (about 10\%).
The result is in good agreement with theoretical predictions.
Rough estimates for two other amplitudes are obtained.
\end{abstract}

\maketitle

\section{Introduction}
 In last years, a lot of experimental and theoretical effort was made
 to study two-photon exchange (TPE) in the elastic electron-proton scattering.
 This activity was motivated by the discovery of the problem
 in the proton form factor measurements: values of form factor ratio $G_E/G_M$
 obtained by Rosenbluth separation and polarization transfer methods 
 were is strong disagreement.
 It is now widely accepted that the discrepancy is caused by TPE,
 but still no direct experimental observations of TPE exist.
 
 Previously, several attempts were made to extract values
 of TPE amplitudes from available experimental data
 in more or less model-independent way \cite{prevArrington,ourPheno}.
 However, lack of precise data and/or theoretical understanding of TPE
 prevented from obtaining sufficiently accurate estimates for the amplitudes.
 In particular, it was difficult to determine the dependence of TPE amplitudes
 on the kinematical parameter $\ve$, since the $\ve$ dependence of polarization
 observables was not known experimentally.
 
 Recently, a search for TPE effects in polarization observables
 was reported \cite{JLab}.
 In this experiment, ratio of transverse and longitudinal
 proton polarization components (polarization ratio)
 was measured with significantly
 improved precision for $Q^2 = 2.5\GeV^2$ and wide range of the parameter $\ve$.

 In the present paper we use latest experimental data to determine TPE
 amplitudes at $Q^2 = 2.5\GeV^2$
 following the ideas of Ref.~\cite{ourPheno}
 with some improvements (described below).
 We will try to obtain as much information on TPE amplitudes as possible,
 while avoiding unnecessary assumptions. 
 In our analysis we only assume that TPE
 is the sole reason for the discrepancy between cross-section
 and polarization data,
 and rely on the following experimental and theoretical facts:
 \begin{enumerate}
  \item The reduced cross-section exhibits no or small nonlinearity in $\ve$
    (\cite{nonlin,Arrington}, also verified in the present work).
  \item The polarization ratio does not vary significantly with $\ve$ \cite{JLab}.
  \item TPE amplitudes must vanish at $\ve\to 1$
    (because they can be represented by convergent dispersion integral;
      $\ve\to 1$ implies $s\to\infty$, where $s$ is c.m. energy squared).
 \end{enumerate}
 The latter point was missing in Ref.~\cite{ourPheno}.

\section{Analysis of cross-section data}
 There were no recent cross-section measurements
 in the $Q^2$ region of our interest, so we have to use older data.
 These data were analyzed before,
 but for our paper to be self-contained
 we repeat such an analysis here.
 We have selected data in the range
 $2.2 \GeV^2 < Q^2 < 2.8 \GeV^2$ \cite{data}.
 The corresponding reduced cross-sections were first multiplied by
 $(1+Q^2/0.71 \GeV^2)^4$, to eliminate most of the $Q^2$ dependence.
 Then, since in Born approximation the reduced cross-section is
\be
 \sigma_R = \tau G_M^2 + \ve G_E^2
\ee
 where $G_E$ and $G_M$ are electric and magnetic form factors,
 $\tau = Q^2/4M^2$ and $M$ is proton mass,
 the resulting values were fitted with the function
\be
 A + B \ve + C (Q^2 - 2.5 \GeV^2)
\ee
 The last term takes into account $Q^2$ dependence of
 $\tau G_M^2$ term in $\sigma_R$.
 We obtain rather acceptable fit with $\chi^2 = 39$ for 28 d.o.f.,
 indicating that the linearity of $\sigma_R$ in $\ve$
 is indeed supported by the data.
 We will mainly need the quantity
\be
 R_{LT}^2 = \tau B/A
\ee
 which would be equal to $(G_E/G_M)^2$ at $Q^2=2.5\GeV^2$ in Born approximation.
 We obtain $R_{LT}^2 = 0.1020 \pm 0.0057$, in agreement
 with the results Ref.~\cite{Arrington} (0.1015).
 In further calculations we use the first value.

\section{Extraction of TPE amplitudes} 
\begin{figure}
 \includegraphics[width=0.48\textwidth]{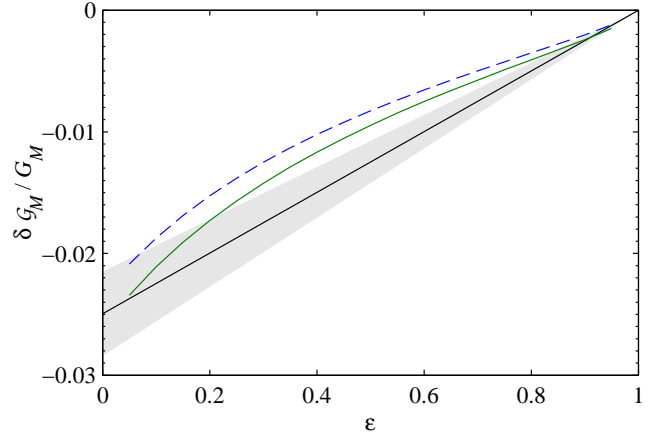}
 \caption{Extracted TPE amplitude $\delta\CG_M/G_M$ (grey, $1\sigma$ band)
   and its theoretical estimates, calculated with:
   elastic intermediate state only \cite{ourDisp} (dashed),
   elastic + $\Delta$ resonance \cite{ourDelta} (solid).}
 \label{Fig1}
\end{figure}

 We will use mostly the same notation as in Ref.~\cite{ourPheno}.
 We denote particle momenta according to
\be
 e(k)+p(p)\to e(k')+p(p'),
\ee
 and define $q = p'-p$, $P=(p+p')/2$, $K=(k+k')/2$, $Q^2 = -q^2$. 
 In presence of TPE, elastic electron-proton
 scattering amplitude has the form
\be
\begin{split}
 {\cal M} = &\frac{4\pi\alpha}{Q^2} \bar u'\gamma_\mu u \,
 \bar U' \left(\tilde F_1 \gamma^\mu - \tilde F_2 
 [\gamma^\mu,\gamma^\nu] \frac{q_\nu}{4M} +\right.\\
 &\left.+\tilde F_3  K_\nu \gamma^\nu \frac{P^\mu}{M^2}
 \right) U
\end{split}
\ee
 where $\alpha$ is fine structure constant, $u$, $u'$ ($U$, $U'$)
 are initial and final electron (proton) spinors,
 and $\tilde F_i$ are scalar invariant amplitudes.
 It is convenient to introduce linear combinations \cite{ourPheno}
\be
 \begin{array}{rclcl}
  \CG_E &=& \tilde F_1 - \tau \tilde F_2 + \nu \tilde F_3/4M^2 &=& G_E + \delta\CG_E\\
  \CG_M &=& \tilde F_1 +\tilde F_2 + \ve \nu \tilde F_3/4M^2   &=& G_M + \delta\CG_M\\
  \CG_3 &=& \nu \tilde F_3/4M^2                                &=& \delta\CG_3
 \end{array}
\ee
 where $\nu = 4PK$ and prefix $\delta$ indicates TPE contribution.
 The TPE amplitudes $\delta\CG_i$ are complex,
 but only their real parts contribute to the observables discussed here.
 Everywhere below, speaking of the amplitudes, we will mean their real parts.
 Neglecting terms of order $\alpha^2$,
 the reduced cross-section and polarization ratio can be written as
\bea
 \sigma_R &=& G_M^2 \left\{ \tau + \ve R_0^2
    + 2\tau \frac{\delta\CG_M}{G_M} 
    + 2\ve R_0^2 \frac{\delta\CG_E}{G_E} \right\}
  \label{sigma_R} \\
 R &=& R_0 \left\{ 1 + \frac{\delta\CG_E}{G_E}
    - \frac{\delta\CG_M}{G_M}
    - \frac{\ve(1-\ve)}{1+\ve} \frac{\delta\CG_3}{G_M} \right\}
  \label{R}
\eea
 where $R_0 = G_E/G_M$. Note that our definition of $R$
 does not include a factor of $\mu \approx 2.793$,
 thus $R$ (and $R_0$) is rather small quantity ($\approx 0.25$ for $Q^2 = 2.5 \GeV^2$).
 Utilizing this fact we will neglect last term in Eq.(\ref{sigma_R})
 (the validity of this approximation will be checked afterwards).
 Then, as it was argued in Ref.~\cite{ourPheno},
 the observed cross-section linearity in $\ve$ {\it forces} us
 to parameterize TPE amplitude $\delta\CG_M$ as a linear function of $\ve$.
 To vanish at $\ve\to 1$, it must have the form
\be \label{CG_M}
 \delta\CG_M/G_M = a(1-\ve)
\ee
 Then we have
\be
 \sigma_R = G_M^2 \{ \tau + \ve R_0^2 + 2 \tau a (1 - \ve) \}
\ee
 and the cross-section slope is
\be
 R_{LT}^2 = \frac{R_0^2 - 2\tau a}{\tau(1+2a)}
\ee
 from which we obtain
\be \label{a}
 a = \frac{R_0^2-R_{LT}^2}{2(\tau+R_{LT}^2)}
\ee
 Together with Eq.(\ref{CG_M}), this fully determines
 the amplitude $\delta\CG_M$.
 As a first approximation, we replace $R_0^2$
 by experimental value of polarization ratio, $R = 0.6923 \pm 0.0058$ \cite{JLab},
 and obtain numerically
\be \label{aNum}
 a = -0.0250 \pm 0.0035
\ee
 Thus extracted amplitude $\delta\CG_M/G_M$ is shown in Fig.~\ref{Fig1}
 with $1\sigma$ error band.
 The theoretical prediction \cite{ourDisp,ourDelta} is also shown and
 agrees rather well with our result.

 Now we will take a closer look on the polarization ratio $R$,
 which allows us to get some information about the amplitude $\delta\CG_E$.
 First, we note that the last term in Eq.(\ref{R}) should be very small,
 because the factor $\frac{\ve(1-\ve)}{1+\ve}$
 is not greater than 0.18 for $0<\ve<1$.
 Thus we are left with
\be \label{R2}
 R = R_0 \left\{ 1 + \frac{\delta\CG_E}{G_E}
    - \frac{\delta\CG_M}{G_M} \right\} 
\ee
 Since both $\delta\CG_M$ and $\delta\CG_E$
 vanish at $\ve\to 1$, we obviously have
\be
 R_0 = \left. R \right|_{\ve=1}, \qquad
 \delta\CG_E - R_0 \delta\CG_M = R - R_0
\ee 
 The experiment says that, at $Q^2=2.5\GeV^2$,
 there is no significant variation of $R$ with $\ve$ \cite{JLab}.
 This implies
\be \label{CG_E}
 R \approx R_0 \qquad {\rm\ and\ }\qquad
 \delta\CG_E \approx R_0 \delta\CG_M 
\ee
 (which also justifies using $R$ instead of $R_0$ in Eq.(\ref{a})).
 Now we can cross-check that the amplitude $\delta\CG_E$
 has small impact on the cross-section.
 Using Eqs.(\ref{CG_M},\ref{a},\ref{CG_E}), we calculate
 the corresponding correction (the last term of Eq.(\ref{sigma_R}))
 and subtract it from the data.
 Then we repeat cross-section fitting and extraction of $\delta\CG_M$,
 as described above. We obtain $a = -0.0248$, i.e. practically no change
 with respect to (\ref{aNum}).
 
\begin{figure}
 \includegraphics[width=0.48\textwidth]{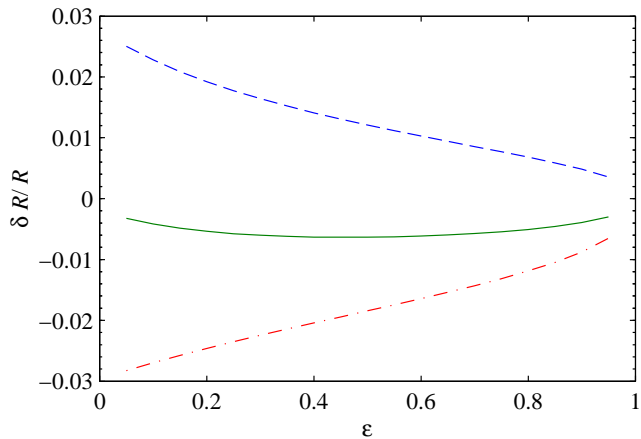}
  \caption{TPE correction to polarization ratio $\delta R/R$, 
    contribution of proton (dashed), $\Delta$ resonance (dash-dotted),
    and total (solid). }
 \label{Fig2}
\end{figure}

 Eqs.(\ref{CG_E}) are, of course, a rough estimate.
 In particular, they do not allow to estimate
 the uncertainty of $\delta\CG_E$.
 An accurate $\CG_E$ extraction with estimation of uncertainties
 requires determination of the small quantity $\delta R = R-R_0$,
 for which the present data hardly suffice.
%
%
 Nevertheless, note that theoretical calculations also show
 relative smallness of $\delta R$,
 which arises from significant cancellation between
 proton and $\Delta$ resonance contributions (Fig.~\ref{Fig2}).

 In the experiment \cite{JLab}, one more quantity was measured:
 longitudinal polarization of the final proton, $P_l$.
 In principle, this data could help to determine
 the remaining amplitude $\delta\CG_3$.
 
\begin{figure}
 \includegraphics[width=0.48\textwidth]{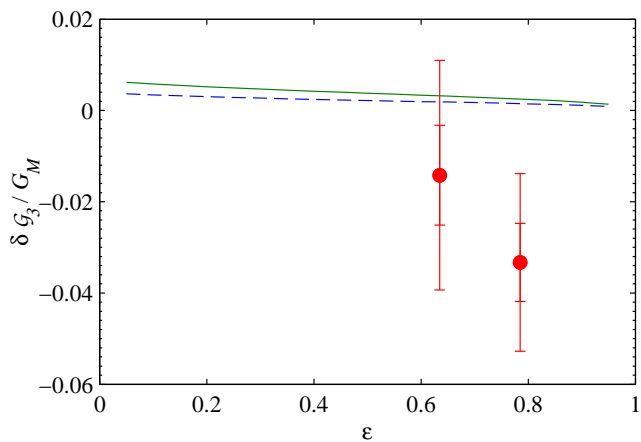}
 \caption{Extracted amplitude $\delta\CG_3/G_M$.
   Statistical and systematic errors are added in quadrature,
   inner bars --- pure statistical errors.
   Theoretical calculations with: elastic intermediate state only \cite{ourDisp} (dashed),
   elastic + $\Delta$ resonance \cite{ourDelta} (solid).}
 \label{Fig3}
\end{figure}

 The TPE correction to $P_l$ is given by
\be
 \delta P_l = - 2\ve P_l \left\{
   \frac{R_0^2 \delta R}{\ve R_0^2 + \tau}
   + \frac{\ve}{1+\ve} \frac{\delta\CG_3}{G_M}
 \right\}
\ee
 where $\delta R = R - R_0$ is TPE correction to polarization ratio
 according to Eq.(\ref{R2}).
 As $R_0^2 \ll 1$, the first term is negligible,
 and the deviation of $P_l$ from its Born value
 is governed by the amplitude $\delta\CG_3$.
 However, two available data points are too few to make any statements
 about $\ve$ dependence of $\delta\CG_3$.
 We can only compute $\delta\CG_3$ at the $\ve$ values of experimental data.
 The results are shown in Fig.~\ref{Fig3}.
 Obviously, their precision is insufficient to make
 a meaningful comparison with theory.
 The obtained values are compatible with zero,
 and do not contradict theoretical estimates as well.

\section{Conclusions}
In summary, we have tried to extract TPE amplitudes for the elastic
electron-proton scattering at $Q^2=2.5\GeV^2$
solely from the experimental data on cross-sections
and polarization observables.
Having defined three independent amplitudes
$\delta\CG_M$, $\delta\CG_E$ and $\delta\CG_3$,
we found that the effect of these amplitudes on the observables is ``decoupled'':
the cross-section is mainly influenced by $\delta\CG_M$,
the polarization ratio --- by $\delta\CG_E$ and $\delta\CG_M$, 
longitudinal polarization component --- by $\delta\CG_3$.
The amplitude $\delta\CG_M$ can be extracted with approximately 10\%
accuracy and is in good agreement with theoretical calculations.
The weakness of polarization ratio variation with $\ve$
implies approximate equality $\delta\CG_E/G_E \approx \delta\CG_M/G_M$.
As to the amplitude $\delta\CG_3$, present experimental data are consistent with $\delta\CG_3 = 0$.
To allow for more accurate extraction of $\delta\CG_E$ and $\delta\CG_3$,
further polarization measurements at different $\ve$ are clearly needed.
It is also worth noting that momentum transfer value $Q^2 = 2.5\GeV^2$
was not good choice for the experiment \cite{JLab}:
just in this region TPE correction to polarization ratio
is especially small because elastic and $\Delta$ resonance contributions
almost cancel each other.

Recently, a preprint \cite{Kivel} appeared,
in which the same problem was considered.
However, authors have used certain parameterization
of polarization component $P_l$, which,
to our opinion, is not well motivated by experimental data.
Their results strongly differ from ours
as well as from theoretical calculations.


\begin{thebibliography}{20}
 \bibitem{prevArrington} J.~Arrington, Phys.Rev. C {\bf 71}, 015202 (2005).
 \bibitem{ourPheno} D.~Borisyuk, A.~Kobushkin, Phys. Rev. C {\bf 76}, 022201(R) (2007).
 \bibitem{JLab} M.~Meziane {\it et al.}, arXiv:1012.0339 [nucl-ex].
 \bibitem{nonlin} V.~Tvaskis {\it et al.}, Phys.~Rev.~C {\bf 73}, 025206 (2006).
 \bibitem{Arrington} J.~Arrington, W.~Melnitchouk, J.A.~Tjon, Phys. Rev. C {\bf 76}, 035205 (2007).
 \bibitem{data}
   J. Litt {\it et al.}, Phys.~Lett.~B {\bf 31}, 47 (1970);
   R.C.~Walker {\it et al.}, Phys.~Rev.~D {\bf 49}, 5671 (1994);
   L.~Andivahis {\it et al.}, Phys.~Rev.~D {\bf 50}, 5491 (1994); 
   M.E.~Christy {\it et al.}, Phys.~Rev.~C {\bf 70}, 015206 (2004);
   I.A.~Qattan, arXiv:nucl-ex/0610006.
 \bibitem{ourDisp} D.~Borisyuk, A.~Kobushkin, Phys. Rev. C {\bf 78}, 025208 (2008).
 \bibitem{ourDelta} D.~Borisyuk, A.~Kobushkin, in preparation.
 \bibitem{Kivel} J.~Guttmann, N.~Kivel, M.~Meziane, M.~Vanderhaeghen, arXiv:1012.0564 [hep-ph].
\end{thebibliography}
\end{document}